# Long anisotropic absolute refractory periods with rapid rise-times to reliable responsiveness

Shira Sardi[1,†], Roni Vardi[2,†], Yael Tugendhaft[1], Anton Sheinin[3], Amir Goldental[1] & Ido Kanter[1,2,*]
[1]Department of Physics, Bar-Ilan University, Ramat-Gan, 52900, Israel.
[2]Gonda Interdisciplinary Brain Research Center and the Goodman Faculty of Life Sciences, Bar-Ilan University, Ramat-Gan, 52900, Israel.
[3]Sagol School of Neuroscience, Tel Aviv University, Tel Aviv, Israel.
[†]These authors contributed equally to this work.
[*]e-mail: ido.kanter@biu.ac.il

Refractoriness is a fundamental property of excitable elements, such as neurons, indicating the probability for re-excitation in a given time-lag, and is typically linked to the neuronal hyperpolarization following an evoked spike. Here we measured the refractory periods (RPs) in neuronal cultures and observed that an average anisotropic absolute RP could exceed 10 milliseconds and its tail 20 milliseconds, independent of a large stimulation frequency range. It is an order of magnitude longer than anticipated and comparable with the decaying membrane potential timescale. It is followed by a sharp rise-time (relative RP) of merely ~1 millisecond to complete responsiveness. Extracellular stimulations result in longer absolute RPs than solely intracellular ones, and a pair of extracellular stimulations from two different routes exhibits distinct absolute RPs, depending on their order. Our results indicate that a neuron is an accurate excitable element, where the diverse RPs cannot be attributed solely to the soma and imply fast mutual interactions between different stimulation routes and dendrites. Further elucidation of neuronal computational capabilities and their interplay with adaptation mechanisms is warranted.

## I. INTRODUCTION

Refractory periods (RPs) are an unavoidable feature of excitable elements, given that following excitation, the formation of necessary conditions for a re-excitation demands a recovery period. This universal feature governs the activity of excitable elements ranging from stimulated and spontaneous emission from atoms[1] to spiking chaotic lasers[2].

The responsiveness of a neuron for a short time re-excitation for a given stimulation amplitude can be classified by the following three consecutive time-lags; an absolute refractory period (ARP) with vanishing spike response probability[3,4], a relative refractory period (RRP) with increasing spike response probability towards one and a complete response probability[4,5]. The ARP is the shortest period and is estimated to last for about 2 milliseconds after an evoked spike. The RRP typically lasts an order of magnitude longer (tens of milliseconds), and its duration and response probability depends on the amplitude of stimulation[3,5]. Hence, there is a significant time-lag where the neuron functions as an unreliable element with stochastic responses[4-6]. This feature is typically attributed to the hyperpolarization of a neuron following an evoked spike[7], which makes it difficult to generate a consecutive evoke spike[8]. Each neuron can be characterized by its scalar ARP and RRP, reproducible for the same stimulation scheduling and amplitude.

This study presents advanced experiments on neuronal cultures without visible hyperpolarization[7,9] and with vanishing or negligible RRP, indicating that the neuron is a reliable threshold unit.The ARP duration is longer than was assumed and can exceed 20 ms, a comparable time-lag to the decay time constant of the neuronal membrane[10,11]. The ARP differs for intra- and extra-cellular stimulations and pairs of extracellular stimulations via different routes or dendrites, indicating that the ARP cannot be attributed solely to the soma.

## II. RESULTS

Our experimental setup consisted of neuronal cultures on a multi-electrode-array combined with a patch-clamp, with added synaptic blockers. Each patched neuron was stimulated intracellularly at the soma or extracellularly via its dendrites[12] (Fig. 1(a) and

Methods) where the recorded spike waveforms do not exhibit visible hyperpolarization (Figs. 1(b)-1(c)).

The first type of experiments consisted of pairs of intracellular stimulations separated by $\Delta t$ (Fig. 1(d)). The pairs were repeated at 0.5 Hz with increasing predefined scheduling for $\Delta t$. The ARP was defined as the shortest time-lag between a pair of evoked spikes. The RRP was the shortest time-lag between a pair of evoked spikes from which they were generated with probability 1. Note that the time-lag between stimulations and their corresponding evoked spikes is practically the same since the neuronal response latency is stable at low stimulation frequency. This type of experiment with stimulation amplitudes close to the neuronal intracellular threshold results in sharp transitions from ARP to a phase with complete responses, indicating a vanishing RRP $\leq \Delta t$ (Fig. 1(e)). The duration of the ARP varies widely between neurons and is measured to be greater than 2 ms, potentially exceeding 8 ms, n=14 mean=5.8 ms, and std=2.4 ms (Fig. 1(f) and Supplemental text). For enhanced stimulation amplitudes, considerably higher than the threshold (Methods), the ARP of the presented neuron is reduced by 2.3 ms (Fig. 1(g)) and with vanishing RRP. One can see that in this example, the ARP is only slightly affected by the intracellular stimulation amplitude.

The second type of experiments consisted of an intracellular stimulation followed by an extracellular stimulation after $\Delta t$, where these pairs were repeated at 0.5 Hz and with a predefined increasing or decreasing scheduling for $\Delta t$ (Fig. 2(a)). The entire neuronal low stimulation frequency, 1 Hz, excludes response failures due to approaching the intermittent phase, together with a stable neuronal response latency necessary to control the relative timings of extracellular stimulations precisely [13]. Results indicate a greater ARP (n=18, mean=12.7 ms and std= 6.4 ms) (Fig. 2(b-e)) than the one obtained in the first type of experiments and with an average 1 ms RRP  (0.5 ms in Fig. 2(c)). Results for another neuron indicated an ARP that exceeded 20 ms (Fig. 2(d)), without visible RRP. We note that the APR greater than 20 ms is not a rare event (Fig. 1(e)). .

The third type of experiment was similar to the second but consisted of a pair of extracellular stimulations from the same electrode (Fig. 2(f)). In this case, the ARP exceeded 10 ms  (n=11, mean= 7.3 ms, std = 3 ms) and the RRP lasted for 1.3 ms  (Figs.

2(g)-2(i)) and 1 ms on the average. These results indicate that the ARP with vanishing RRP is a phenomenon that is quantitatively independent for a large range of stimulation frequencies (Fig. 3 and Supplementary fig. S1).

The fourth type of experiments consisted of a pair of extracellular stimulations separated by $\Delta t$ (Fig. 4(a)), similar to the third type of experiments but employing *different* extracellular electrodes. The two extracellular stimulations (green and purple in Fig. 4(a), where a different color indicates a different extracellular stimulating electrode) generated different spike waveforms (Fig. 4(b)), indicating that they involved two different dendrites, where the neuron functions as a multiple threshold unit[9]. In one of the spike waveforms, the membrane potential decayed monotonically after an evoked spike (Fig. 4(b), purple). In contrast, the second spike waveform exhibited a momentary depolarization after the evoked spike (Fig. 4(b), green). In the presented experiment, for pairs of "green-purple" stimulations, the ARP resulted in 17 ms (Fig. 4(c)), while for pairs of "purple-green" stimulations, the ARP resulted in only 8 ms (Fig. 4(d)). Note that for both cases, the RRP vanished under the resolution of $\Delta t = 0.5$ ms, where both spike waveforms did not exhibit visible hyperpolarization. Our results indicate that different stimulation routes have different ARPs; hence, the phenomenon of the RP cannot be attributed solely to the soma[14,15]. Note that the variations in the ARP within each stimulation route (Supplementary fig. S2) is less than the difference between their ARPs. Nevertheless, the possible mutual interaction between the two different stimulation routes demands further examination.

The anisotropic property of the ARP was also tested for two extracellular electrodes that generated similar spike waveforms (Fig. 5). This reality in our experiments indicates stimulation either via two branches of the same dendrite or via two dendrites generating the same spike waveform. All four different orders of a pair of stimulations via the two extracellular electrodes are presented (Fig. 5). Stimulation scheduling via each one of the electrodes resulted in a very similar ARP (Figs. 5(a)-5(b)), and with vanishing RRP. However, an increase in the ARP (beyond fluctuations, Supplementary fig. S2) was examined when stimulation scheduling involved two different stimulation sources (Figs. 5(c)-5(d)), and with vanishing RRP. The distinct ARP for a different order of extracellular stimulations indicates that the ARP duration is not solely a feature of the soma but is strongly affected by the stimulation route. Moreover, there was a mutual interaction between the two stimulation routes. For instance, the ARP for a pair of purple-green stimulations (Fig. 5(c)) differs from the ARP of green-green stimulations (Fig. 5(a)) and the ARP of green-purple stimulations (Fig. 5(d)). Limited experimental results (n=4) indicate that the ARPs measured from different stimulation routes (26, 20.8, 23, 16 ms) are longer than ARPs measured from the same stimulation route (12, 12, 10, 10 ms) indicating an average increase of 10.5 ms. This increase is greater than the fluctuations of repeated measurements using the same electrode, less than 1 ms (Supplementary fig. S2). Note that the longer ARPs measured from different stimulation routes (Figs. 5(c)-5(d)) in comparison to a pair of stimulations via the same route (Figs. 5(a)-5(b)), might indicate a repulsion mechanism as a result of the neural backpropagation to the dendrite.

The distinct ARP for different routes (Figs. 4, 5) was also measured by comparing pairs of intercellular stimulations (Fig. 1(d)) and pairs of intra- and extra-cellular stimulations (Fig. 2(a)). While pairs of strong intracellular stimulations (Methods) resulted in 4.8 ms ARP (Fig. 6(a)), pairs of strong intra- followed by extra-cellular stimulations using the same neuron presented much longer ARP, 13.5 ms (Fig. 6(b)), both with a vanishing RRP. Pairs of extra- and intra-cellular stimulations resulted in 6 ms ARP (Fig. 6(c)), while its duration for the opposite order of pairs of the same neuron was 8.5 ms (Fig. 2(b)). These preliminary results indicate that an intracellular evoked spike affected the ARP of an extracellular stimulation.

Reported phenomena are based on more than 50 examined cultures, where each type of experiment was repeated about 10 times. Note that no hyperpolarization was observed in all examined neurons. The experimental measured RPs were found to cover a large range. In addition, the reported range of the RPs is based on different types of neurons representing a variety of spike waveforms. Results do not allow deducing the distribution of the RPs and preferred ranges. For instance, for experimental type 1, the ARP was found to be in the range of [2, 10] ms; for experimental type 2, the ARP was found to be in the range of [5, 30] ms; for experimental type 3, the ARP was found to be in the range of [3, 13] ms. See Supplementary text for detailed results.

## III. CONCLUSION

This study's results raise the following fundamental questions regarding the underlying mechanisms governing the dynamics of the neuron: (a) The mechanism governing the sharp transition between ARP and neuronal complete responsiveness; (b) The interplay between quantitative dendritic features and the phenomenon anisotropic ARP, and whether the anisotropic ARP feature is attributed to a dendrite or its branches; (c) The source for mutual interactions between different dendrites affecting their ARPs; (d) A possible interplay between neuronal temporal and spatial summation mechanism and the ARP, where both quantities could have similar timescales. Direct answers to these questions and identifying the underlying mechanisms require higher experimental resolutions[14,16,17], beyond our experimental setup. However, in the discussion below, we elaborated on some of their possible aspects.

The practically vanishing RRPs compared to the duration of the ARPs, for neurons without visible hyperpolarization is consistent with the traditional assumption regarding the interrelation between these two dynamical features, RRP, and hyperpolarization. The effect of hyperpolarization has a stochastic nature when the neuron is repeatedly stimulated[18,19]; hence, stochastic responses are expected to result in RRP. In addition, the hyperpolarization phenomenon is expected to extend the RRP further. Nevertheless, one cannot exclude the possibility that the examined neurons will exhibit hyperpolarization in different stimulation scenarios, amplitude, duration, and the use of various extracellular solutions. In such a case, one might expect RRP re-appearance, or the interrelation between RRP and hyperpolarization will be in question. We noted that the reported range of the RPs based on many neurons, without hyperpolarization, representing a variety of spike waveforms are limited. It does not allow deducing the RP distribution and the preferred ranges. These questions, and the robustness of RPs to temperature variations[20], are out of the scope of our study and deserve further detailed research.

The consistent time dilation in the ARP in some of the scenarios where extracellular stimulations are involved (Figs 4, 5), compared to sole pairs of intracellular stimulations, strongly indicates that dendrites contribute to the recovery time of re-excitation. Hence, the ARP in the activity of neural networks is a function of the soma and dendrites features. Results also indicated that a spike generated by one route affects the ARP of another (Figs. 4, 5). The mechanism underlying the anisotropic nature of the neuron is unknown; however, it could be explained by different populations of calcium, sodium, and potassium channels[14] in the different parts of dendritic arbors[21-23]. This mutual effect among stimulation routes is within dendritic adaptation[12], where a spike generated by one dendrite changes the strength of another dendrite, which generates a momentary depolarization.

A neuron with an absence of RRP, or its minimal duration, functions as a reliable spiking threshold unit with a negligible rise-time. However, the dynamic significance of long ARPs (Figs. 2-6) is in question, especially where they are comparable with the decaying membrane potential constant (10-20 ms), controlling temporal and spatial summation[10,11]. This strong repulsion among nearby evoked spikes might indicate that spatial summation is effectively prevented as long as a remnant membrane memory is significant. An additional neuronal spike is possible only after the effect of the previous one almost disappears, which is an important ingredient for the enhancement of the signal-to-noise ratio in any communication system.

In some of the experiments, the second stimulation above threshold in a stimulations pair (Fig. 6(d)) showed a momentary depolarization within the ARP. As these stimulations' amplitude decreased toward the threshold, the momentary depolarization appearance time is expected to increase towards the ARP (Fig. 6(d)), while its amplitude remained almost unchanged. This phenomenon indicates a mechanism for mapping stimulation amplitudes to depolarization time-lags[24,25]. It represents a decreased tendency of dendritic adaptation or synaptic adaptation in spike-time-dependent-plasticity[26] with decreasing stimulation amplitudes.

The several ARP durations obtained among various stimulation scenarios of the same neuron call for a better understanding of their possible advanced dendritic computations[14,16,17] and their influence on somatic integration[27-31]. A possible advantage of this phenomenon is exemplified using the theoretical framework of a neuron functioning as a classifier, the perceptron[32-37]. For the simplest neuronal threshold unit, the output for a synchronous input is either a spike or no-spike, regardless of the input amplitudes and the set of weights (Fig. 6($e_1$)). However, for a neuron with two dendrites or two brunches of the same dendrite, represented by different delays (Fig. 6($e_2$)), even with equal weights, the output is more structured due to the effect of long refractory periods. It consists of zero, one, or two spikes and with different possible timings. The more structured output contains partial information about the timings and the amplitudes arriving from the two input branches (pink and green in Fig. 6($e_2$)). For instance, assuming the two branches represent input from different directions, left and right; it is clear that the detailed output, although not uniquely defined, hints the input timings and their strengths (Fig. 6($e_2$)). The mutual information between the output and the input is enhanced using the various properties of the ARPs.

Finally, we note that RP is a universal feature governing the activity of excitable elements ranging from stimulated and spontaneous emission from atoms[38] or spiking chaotic lasers[2] to the repetition of epileptic seizures[39]. It is possible that the understanding of similar phenomena in other scientific areas will be a source for understanding the neuronal RPs.

## IV. MATERIALS AND METHODS

The methods are detailed in [9,12,13].

### Animals

All procedures were in accordance with the National Institutes of Health Guide for the Care and Use of Laboratory Animals and Bar-Ilan University Guidelines for the Use and Care of Laboratory Animals in Research and were approved and supervised by the Bar-Ilan University Animal Care and Use Committee.

### Culture preparation

Cortical neurons were obtained from newborn rats (Sprague-Dawley) within 48 h after birth using mechanical and enzymatic procedures. The cortical tissue was digested enzymatically with 0.05% trypsin solution in phosphate-buffered saline (Dulbecco's PBS) free of calcium and magnesium, and supplemented with 20 mM glucose, at 37°C. Enzyme treatment was terminated using heat-inactivated horse serum, and cells were then mechanically dissociated mostly by trituration. The neurons were plated directly onto substrate-integrated multi-electrode arrays (MEAs) and allowed to develop functionally and structurally mature networks over a time period of 2–4 weeks in vitro, prior to the experiments. The number of plated neurons in a typical network was in the order of 1,300,000, covering an area of about ~5 cm$^2$. The preparations were bathed in minimal essential medium (MEM-Earle, Earle's Salt Base without L-Glutamine) supplemented with heat-inactivated horse serum (5%), B27 supplement (2%), glutamine (0.5mM), glucose (20mM), and gentamicin (10 g/ml), and maintained in an atmosphere of 37°C, 5% $CO_2$ and 95% air in an incubator. All experiments were performed at a temperature of 37°C.

**Synaptic blockers**

Experiments were conducted on cultured cortical neurons that were functionally isolated from their network by a pharmacological block of glutamatergic and GABAergic synapses. For each culture 4–20 μl of a cocktail of synaptic blockers were used, consisting of 10 μM CNQX (6-cyano-7-nitroquinoxaline-2, 3-dione), 80 μM APV (DL-2-amino-5-phosphonovaleric acid) and 5 μM Bicuculline methiodide. This minimal cocktail blocked completely the spontaneous network activity. Blockers were added until no spontaneous activity was observed both in the MEA and in the patch clamp recording. In addition, repeated extracellular stimulations did not provoke the slightest cascades of neuronal responses.

**Stimulation and recording – MEA**

An array of 60 Ti/Au/TiN extracellular electrodes, 30 μm in diameter, and spaced 200 or 500 μm from each other (Multi-Channel Systems, Reutlingen, Germany) was used. The insulation layer (silicon nitride) was pre-treated with polyethyleneimine (0.01% in 0.1 M Borate buffer solution). A commercial setup (MEA2100-60-headstage, MEA2100-interface board, MCS, Reutlingen, Germany) for recording and analyzing data from 60-electrode MEAs was used, with integrated data acquisition from 60 MEA electrodes and 4 additional analog channels, integrated filter amplifier and 3-channel current or voltage stimulus generator. Each channel was sampled at a frequency of 50k samples/s, thus the recorded action potentials and the changes in the neuronal response latency were measured at a resolution of 20 μs. Mono-phasic square voltage pulses were used, in the range of [−900, −100] mV and [100, 2000] μs.

**Stimulation and recording – Patch Clamp**

The Electrophysiological recordings were performed in whole cell configuration utilizing a Multiclamp 700B patch clamp amplifier (Molecular Devices, Foster City, CA). The cells were constantly perfused with the slow flow of extracellular solution consisting of (mM): NaCl 140, KCl 3, CaCl2 2, MgCl2 1, HEPES 10 (Sigma-Aldrich Corp. Rehovot, Israel), supplemented with 2mg/ml glucose (Sigma-Aldrich Corp. Rehovot, Israel), pH 7.3, osmolarity adjusted to 300–305 mOsm. The patch pipettes had resistances of 3–5 MOhm after filling with a solution containing (in mM): KCl 135, HEPES 10, glucose 5, MgATP 2, GTP 0.5 (Sigma-Aldrich Corp. Rehovot, Israel), pH 7.3, osmolarity adjusted to 285–290 mOsm. After obtaining the giga-ohm seal, the membrane was ruptured and the cells were subjected to fast current clamp by injecting an appropriate amount of current in order to adjust the membrane potential to about −70 mV. The changes in neuronal membrane potential were acquired through a Digidata 1550 analog/digital converter using pClamp 10 electrophysiological software (Molecular Devices, Foster City, CA). The acquisition started upon receiving the TTL trigger from MEA setup. The signals were filtered at 10 kHz and digitized at 50 kHz. The cultures mainly consisted of pyramidal cells. For patch clamp recordings, pyramidal neurons were intentionally selected based on their morphological properties.

**MEA and Patch Clamp synchronization**

The experimental setup combines multi-electrode array, MEA 2100, and patch clamp. The multi-electrode array is controlled by the MEA interface boarded and a computer. The Patch clamp sub-system consists of several microstar manipulators, an upright microscope (Slicescope-pro-6000, Scientifca), and a camera. Stimulations and recordings are implemented using multiclamp 700B and Digidata 1550A and are controlled by a second computer. The recorded MEA/patch data is saved on the computers respectively. The time of the MEA system is controlled by a clock placed in the MEA interface board and the time of the patch subsystem is controlled by a clock placed in the Digidata 1550A. The relative timings are controlled by triggers sent from the MEA interface board to the Digidata using leader-laggard configuration.

**Extracellular electrode selection**

For the extracellular stimulations in the performed experiments an extracellular electrode out of the 60 electrodes was chosen by the following procedure. While recording intracellularly, all 60 extracellular electrodes were stimulated serially at 2Hz and above-threshold, where each electrode is stimulated several times. The electrodes that evoked well-isolated, well-formed spikes were used in the experiments. After choosing the extracellular electrode we verify that the stimulation is not antidromic, by verifying that the neuronal response latency is several ms long. In addition, the neuronal response latency is verified to increase by several ms with the number of stimulations. In case the stimulation frequency is above the critical frequency, $f_c$, response failures occur. We report only experiments where the leak of the patched neuron was stable. Note also that the spike waveform remained unchanged during the entire experiment for the same stimulation scheduling.

**Extracellular threshold estimation**

After choosing an extracellular electrode, its threshold for stimulation was estimated. Stimulations at 0.5Hz with duration of 2 ms and different values of voltage were given (using a typical step of 50 mV), until a response failure occurred. The threshold was defined as the average between the stimulation voltage that resulted in a response failure and the closest value of stimulation voltage that resulted in an evoked spike. For patched neurons that were significantly close to an extracellular electrode (several micrometers) shorter stimulation durations were used in order to avoid the stimulation artifact in the voltage recordings.

**Intracellular threshold estimation**

In order to find a threshold for the intracellular stimulation, several stimulations at 1Hz were given. The duration of the stimulations was set to 1-3 milliseconds, and the intensity ranged from 100 pA and increased by 50 pA every stimulation until an evoked spike occurred. The threshold was defined as the average between the stimulation intensity that resulted in a response failure to the closest value of stimulation intensity that resulted in an evoked spike.

**Stimulations strength**

A strong stimulation is a stimulation with an intensity which is at least 15% above the threshold. A stimulation close to the threshold is a stimulation with the lowest intensity found to produce reliable evoked spikes.

**Neuronal response latency**

The neuronal response latency (NRL) is defined as the time-lag between a stimulation pulse onset and its corresponding evoked spike measured by crossing a threshold of -10 mV. In Figs. 2-6 the NRL was estimated in order to adjust the arrival timings of the intra- and extra- cellular stimulations.

**Response probability calculation**

The response probability presented in Figs. 2(c) and 2(g), was calculated as follows. The response to the second stimulation for each $\Delta t$ was marked as 1 for an evoked spike, and 0 for a response failure. We used a sliding window of 3 (Fig. 2(c)) or 5 (Fig. 2(g)) consecutive stimulations on the binary vector and calculated the average response probability for each $\Delta t$. Hence, the response probability in Fig. 2(c) can take the values 0, 1/3, 2/3, 1, and in Fig. 2(g) the values 0, 1/5, 2/5, 3/5, 4/5, 1.

**Data analysis**

Analyses were performed in a Matlab environment (MathWorks, Natwick, MA, USA). The recorded data from the MEA (voltage) was filtered by convolution with a Gaussian that has a standard deviation (STD) of 0.1 ms. Evoked spikes were detected by threshold crossing, -10 mV, using a detection window of 0.5–30 ms following the beginning of an electrical stimulation.

**ACKNOWLEDGMENTS**

This research was supported by the TELEM grant of the Council for Higher Education of Israel.

**AUTHOR CONTRIBUTIONS**

S.S. and R.V. contributed equally to this work. S.S., R.V., and Y.T. prepared the tissue cultures. S.S. and R.V. performed the experiments and analyzed the data. A.G. helped in the operation of the combined MEA and patch system. I.K. initiated the study and supervised all aspects of the work. All authors discussed the results and commented on the manuscript.

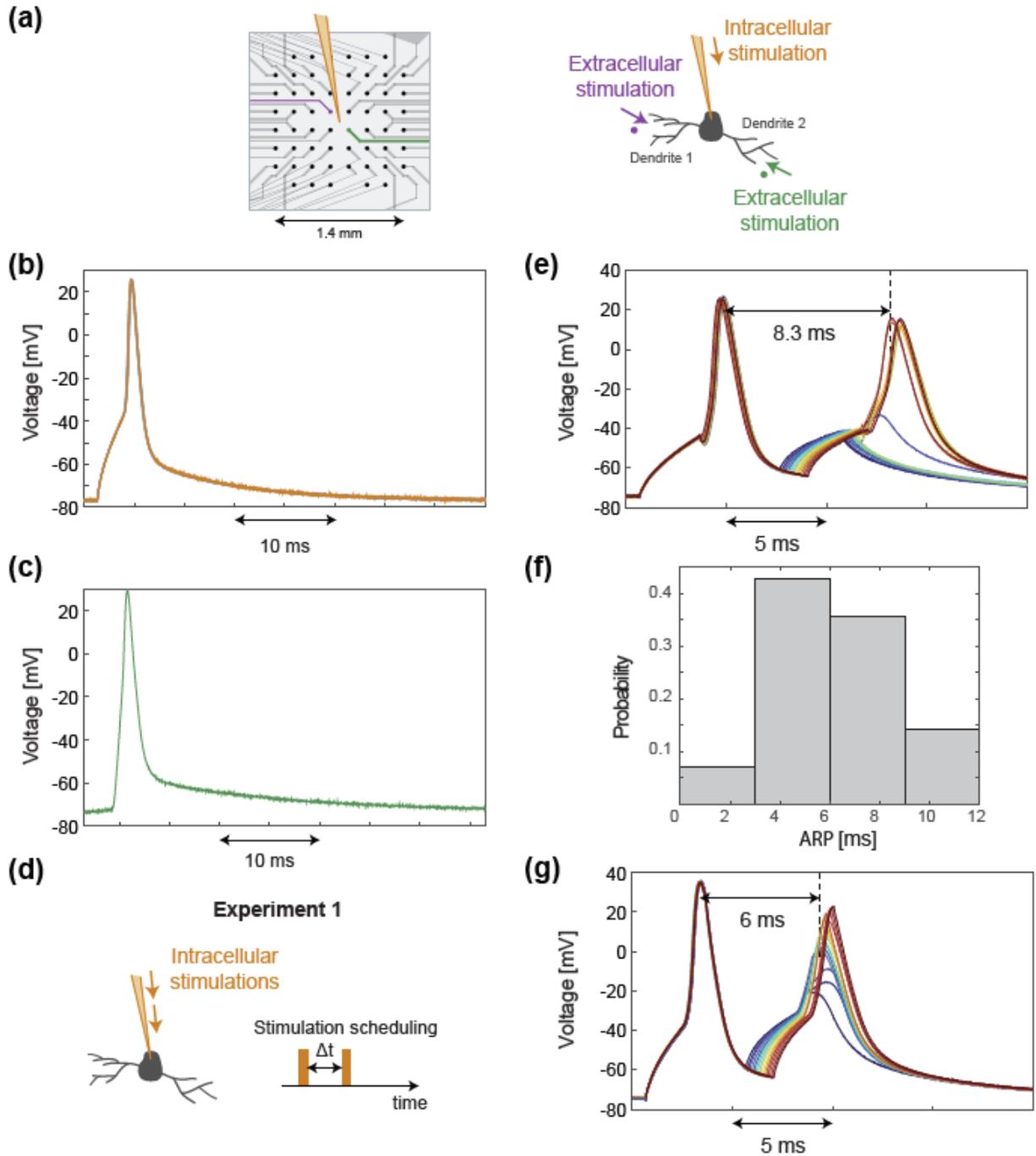

FIG. 1. The experimental setup and measurements of the RP using pairs of intracellular stimulations. (a) Left: A micro-electrode array (MEA) consisting of 60 electrodes combined with a patch-clamp pipette (orange) (see Methods for details). Right: A scheme of a patched neuron, which can be extracellularly stimulated via its two dendrites (purple and green electrodes) and intracellularly via the patch-clamp pipette (orange). (b) An intracellular recorded evoked spike by intracellular stimulation without visible hyperpolarization. (c) Similar to (b), but with extracellular

stimulation. (d) A scheme of the first type of experiments where two intracellular stimulations separated by increasing $\Delta t$ is given to a patched neuron. (e) Neuronal responses to the stimulation scheduling in (d) with stimulations amplitudes close to the neuronal intracellular threshold (750 pA intensity, 3 ms duration), where $\Delta t = [4, 5.4]$ ms with increments of 0.1 ms, with vanishing RRP under experimental resolution. (f) Histogram of the ARP obtained in n=14 experiments as in (e) (see also Supplemental text). (g) The same neuron in (e) but with strong intracellular stimulations amplitudes (1200 pA intensity, 3 ms duration), $\Delta t = [3, 6.8]$ ms with increments of 0.2 ms, with vanishing RRP under experimental resolution.

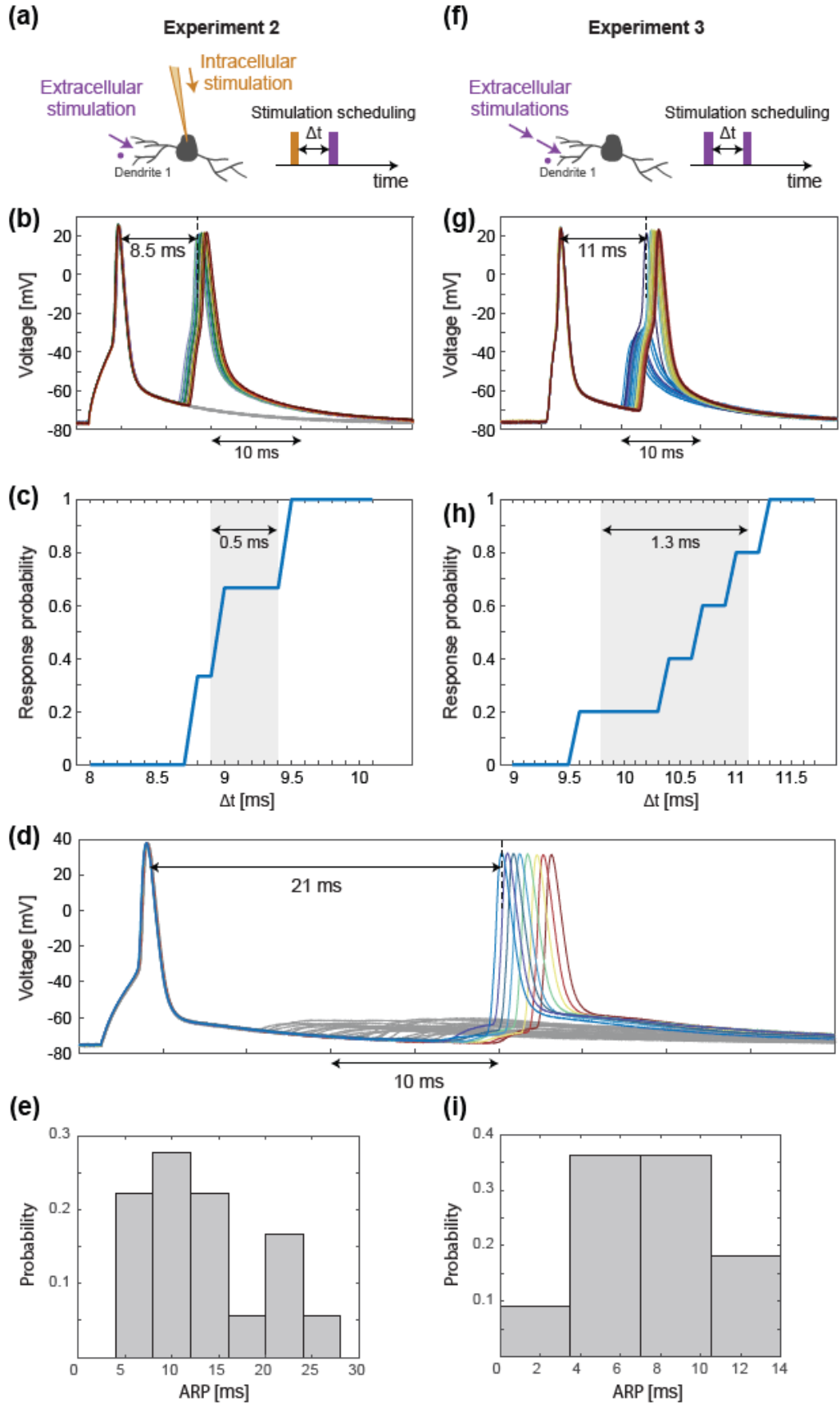

FIG. 2. Measurements of the RP using combinations of intra- and extra-cellular stimulations. (a) A scheme of the second type of experiments where a neuron is stimulated intracellularly and then stimulated extracellularly after $\Delta t$. (b) Results of (a) using $\Delta t = [8, 12.9]$ ms with an increment of 0.1 ms (gray color lines represent no responses to the second extracellular stimulations, as the neuron is in the ARP). Extracellular stimulations were given with 380 mV amplitude, 0.2 ms duration. Intracellular stimulations were given with 1200 pA intensity, 3 ms duration. (c) Probability for an evoked spike in (b) using sliding average over 3 consecutive $\Delta t$ (the responses in the range [8.9,9.4] are 101101, where 1/0 stands for spike/no-spike, respectively), RRP= 0.5 ms. (d) Results of (a) for another neuron with $\Delta t = [3, 27.5]$ ms with increments of 0.5 ms, indicating ARP larger than 20 ms and negligible RRP (gray color lines represent small depolarization to the second extracellular stimulations, as the neuron is in the ARP). Extracellular stimulations were given with 900 mV amplitude, 2 ms duration. Intracellular stimulations were given with 800 pA intensity, 3 ms duration. (e) Histogram of ARP of experiments of type 2 (see text and Supplementary text)  (f) A scheme of the third type of experiments where a neuron is stimulated from the same extracellular electrode by pairs of stimulations separated by increasing $\Delta t$. (g) Results of (f) using $\Delta t = [9, 13.9]$ ms with increments of 0.1 ms. Extracellular stimulations were given with 400 mV amplitude, 0.2 ms duration. (h) Probability for an evoked spike in (g) using sliding average over 5 consecutive $\Delta t$ in the range $\Delta t = [9.8, 11.1]$ ms (the responses in the range 10000100101101), RRP= 1.3 ms. (i) Histogram of ARP of experiments of type 3 (see text and Supplementary text).

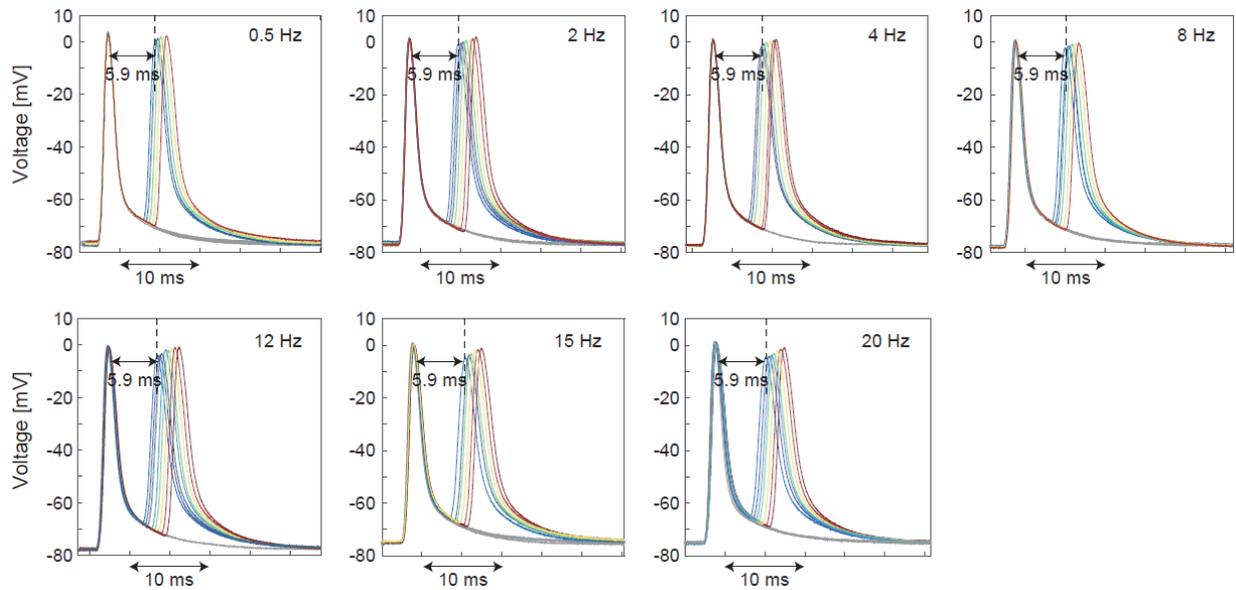

FIG. 3. The ARP as a function of the stimulation frequency, measured by pairs of extracellular stimulations from the same electrode. The patched neuron was stimulated twice using an extracellular electrode, at a time-lag $\Delta t = [3, 27.5]$ ms with increments of 0.5 ms. The neuronal responses are presented for stimulation frequencies of 0.5, 2, 4, 8, 12, 15, and 20 Hz. Results indicate that the ARP is around 5.9 ms and with negligible RRP under the experimental resolution, independent of the stimulation frequency. Gray color lines represent no responses to the second extracellular stimulations, as the neuron is in the ARP. Extracellular stimulations were given with 900 mV amplitude, 0.2 ms duration. Population behavior is presented in Supplementary fig. S1.

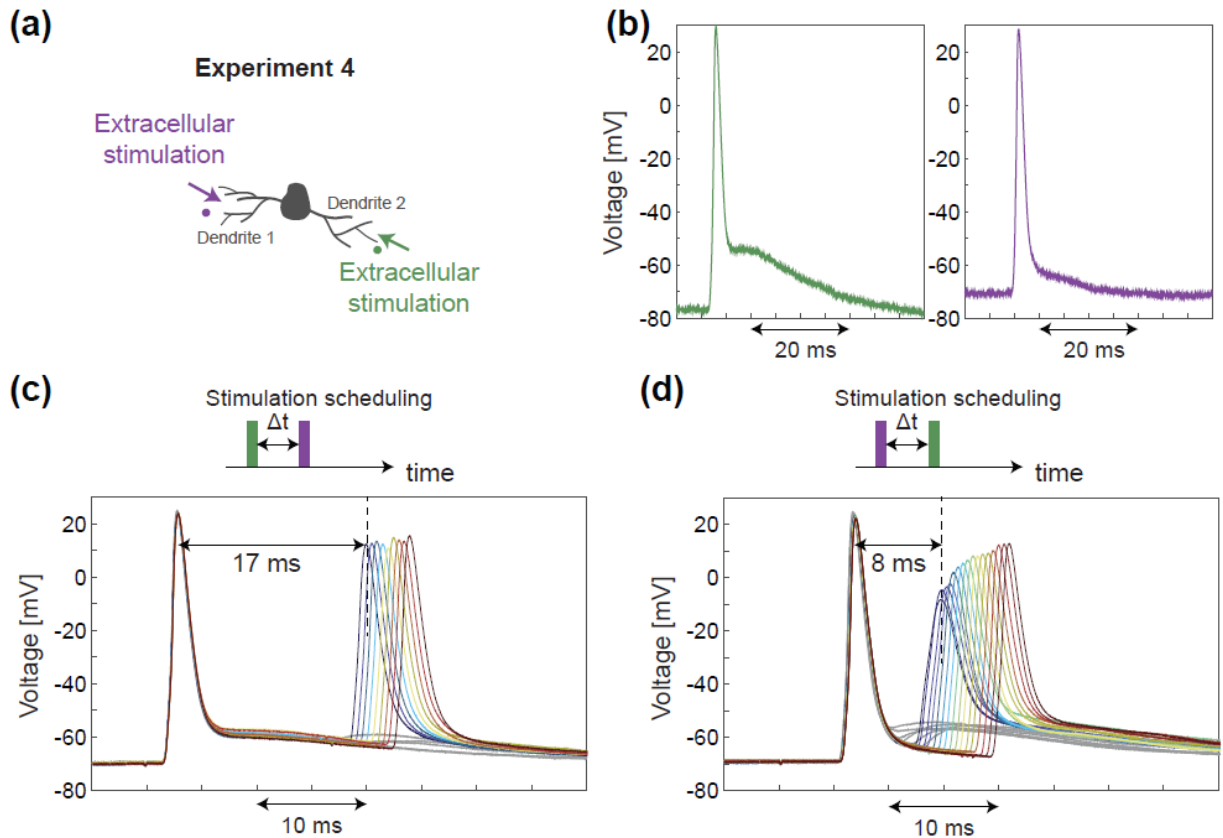

FIG. 4. Anisotropic properties of the ARP, using pairs of extracellular stimulations from two different extracellular electrodes with distinct spike waveforms. (a) A scheme of the fourth type of experiments, a patched neuron is stimulated using two different extracellular electrodes (purple and green) separated by increasing $\Delta t$. (b) The two extracellular electrodes experimentally generate two distinct intracellular measured spike waveforms. (c) Results of (a), where the stimulation from the green electrode is given $\Delta t$ before the purple electrode, $\Delta t = [1,25.5]$ ms with increments of 0.5 ms and vanishing RRP under the experimental resolution (gray color lines represent small depolarization to the second extracellular stimulations, as the neuron is in the ARP). (d) Similar to (c), but the stimulation from the purple electrode is given before the stimulation from the green electrode, $\Delta t = [4,28.5]$ ms with increments of 0.5 ms and vanishing RRP under te expepimental resolution. All extracellular stimulations were given with 900 mV amplitude, 0.2 ms duration.

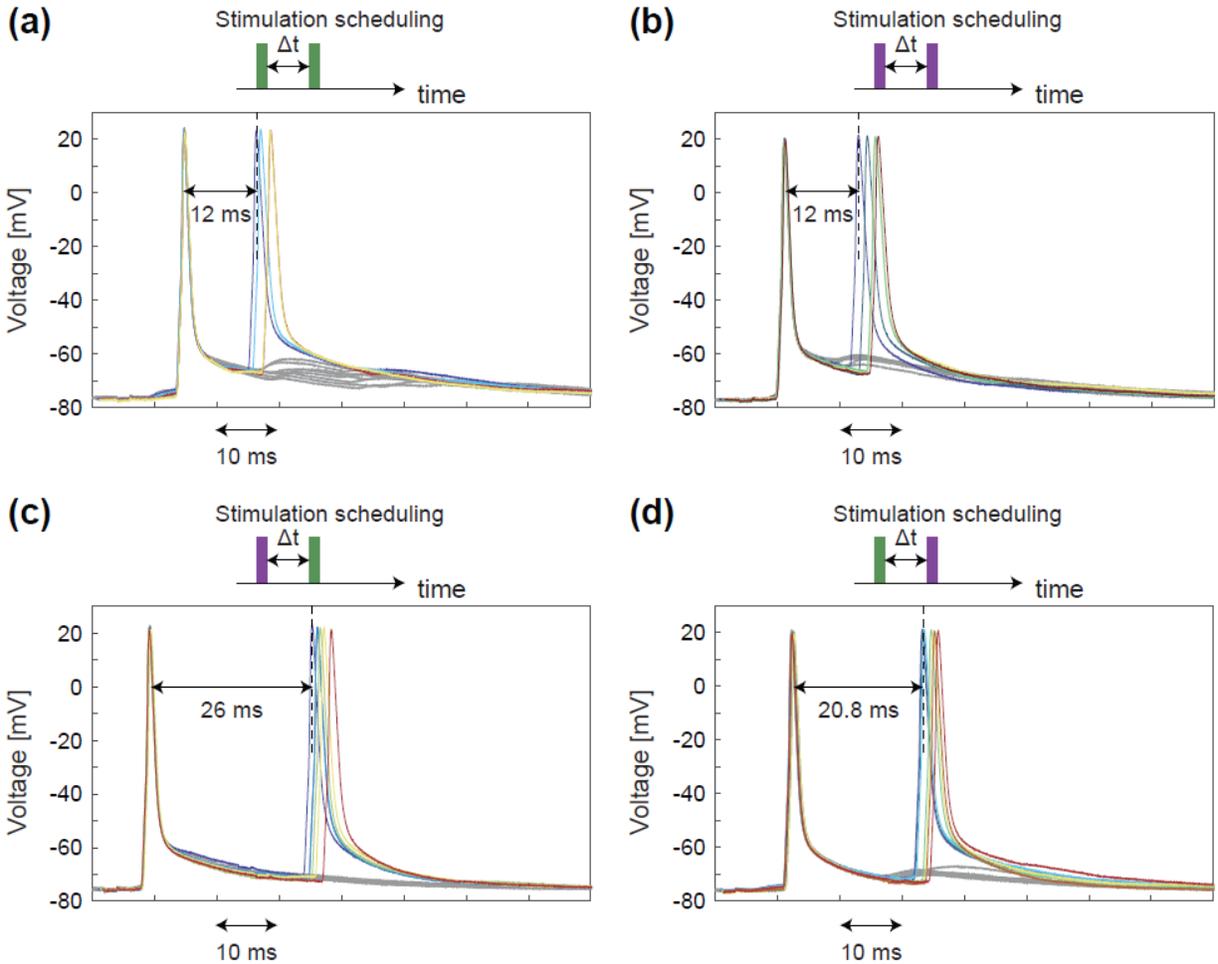

FIG. 5. Anisotropic properties of the ARP, using pairs of extracellular stimulations from two different extracellular electrodes (green and purple) with similar spike waveforms. (a) Top: A scheme of the stimulation scheduling from an extracellular electrode (green), where a pair of stimulations is separated by increasing $\Delta t$. Bottom: The responses of the patched neuron, with vanishing RRP under experimental resolution. (b) Similar to (a), where stimulations are given from the purple electrode and with vanishing RRP under experimental resolution. (c) Similar to (a), where the stimulation from the purple electrode is given before the stimulation from the green electrode, with vanishing RRP under experimental resolution. (d) Similar to (a), the stimulation from the green electrode is given before the stimulation from the purple electrode. In all panels $\Delta t = [2.5, 27]$ ms with increments of 0.5 ms and with vanishing RRP under experimental resolution. Gray color lines represent small depolarization or no responses to the second extracellular stimulations, as the neuron is in the ARP. All extracellular stimulations were given with 900 mV amplitude, 2 ms duration.

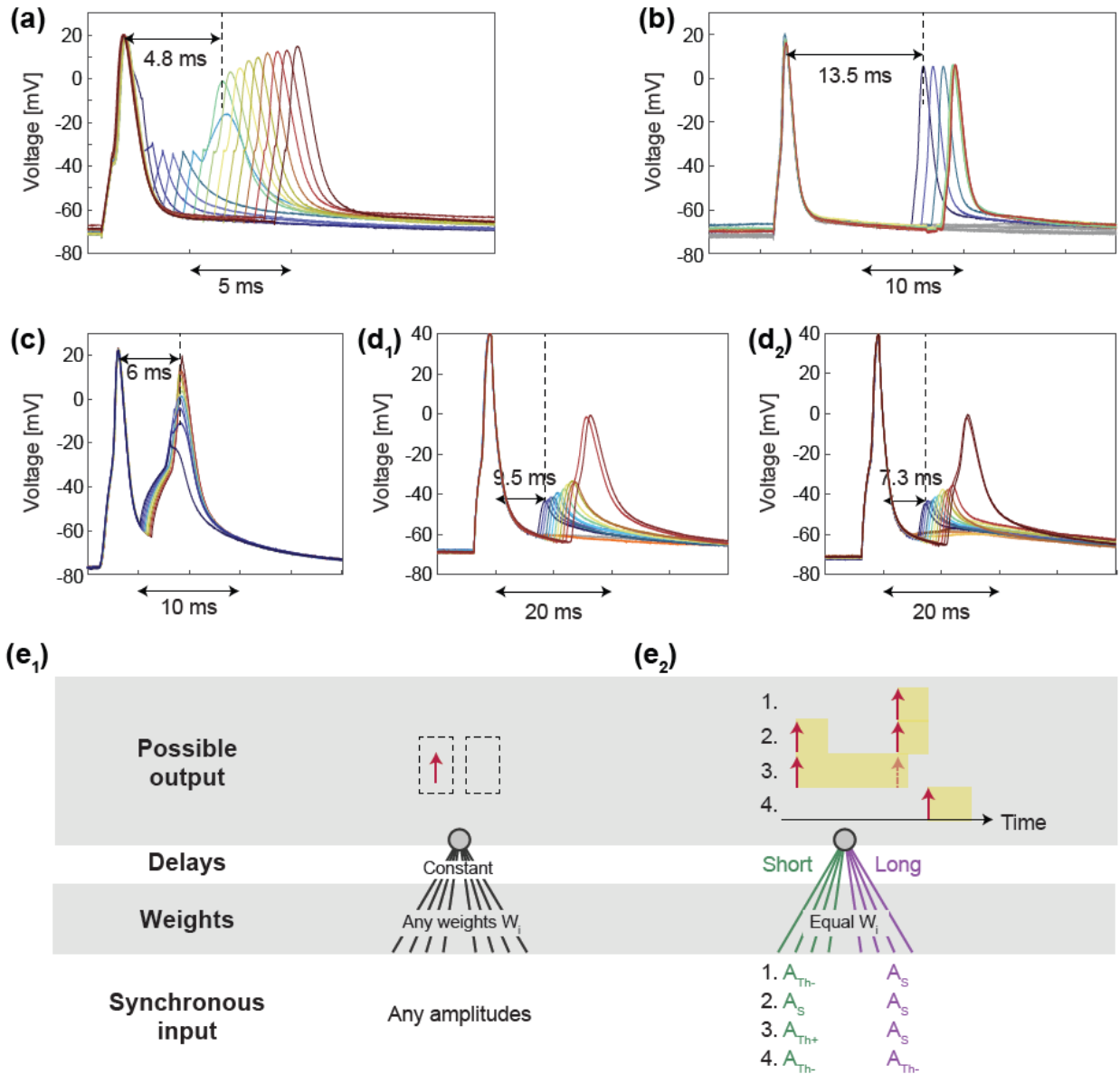

FIG. 6. The effect of different amplitudes and temporal ordering of intra- and extra- cellular stimulations on the ARP and neuronal outputs. (a) Neuronal responses to strong intracellular stimulations amplitudes, where two intracellular stimulations separated by increasing $\Delta t$ are given to a patched neuron (the same type of experiment presented in Fig. 1(d)), where $\Delta t = [1,8]$ ms with increments of 0.5 ms and with vanishing RRP under experimental resolution (gray color lines represent no responses to the second intracellular stimulations, as the neuron is in the ARP). Intracellular stimulations were given with 1300 pA intensity, 0.5 ms duration. (b) Responses of the same neuron in (a) using stimulation scheduling where the neuron is stimulated from the same extracellular electrode by pairs of stimulations separated by increasing $\Delta t$ (the same type of experiment presented in Fig. 2(e)), where $\Delta t = [4, 26]$ ms, with increments of 1 ms and with vanishing RRP under experimental resolution (gray color lines represent no responses to the second extracellular stimulations, as the neuron is in the ARP). Extracellular stimulations were given with 900 mV amplitude, 1 ms duration. (c) The same neuron as in Fig. 2(b), stimulated by pairs of a strong intracellular stimulation given Δt after an extracellular one with an amplitude close to its threshold, $\Delta t = [-6, -1.1]$ ms, with increments of 0.1 ms, where the neuronal response latency of an extracellular stimulation is ~2 ms (Methods) and with vanishing RRP under experimental resolution. Extracellular stimulations were given with 500 mV amplitude, 0.2 ms duration. Intracellular stimulations were given with of 1200 pA intensity, 3 ms duration. (d) The second type of experiment, where the neuron is stimulated by pairs of an extracellular stimulation given Δt after an intracellular one (Fig. 2(a)), with stimulation amplitude of 730 mV (d$_1$) and 900 mV (d$_2$), both durations are 0.2 ms (Methods) and with vanishing RRP under experimental resolution. (e$_1$) A scheme of a perceptron with arbitrary synchronous input and set of weights resulting in two possible outputs, spike or no-spike. (e$_2$) Similar to (e$_1$) but with equal weights which are grouped into short and long delays, representing two dendrites, and with 3 possible stimulation amplitudes (color-coded): $A_s$, strong stimulation much above the threshold, $A_{Th}^+$ and $A_{Th}^-$ close to the threshold from above and below, respectively. Spike timings and ARPs are represented by red-arrows and yellow-strips, respectively. The ARP is longer after an evoked spike resulted from $A_{Th}^+$ (3rd output), where the dashed-red-arrow represents a response failure, no-spike.

# Supplemental Material

# Long anisotropic absolute refractory periods with rapid rise-times to reliable responsiveness

Shira Sardi[1,†], Roni Vardi[2,†], Yael Tugendhaft[1], Anton Sheinin[3], Amir Goldental[1] & Ido Kanter[1,2,*]

[1]Department of Physics, Bar-Ilan University, Ramat-Gan, 52900, Israel.
[2]Gonda Interdisciplinary Brain Research Center and the Goodman Faculty of Life Sciences, Bar-Ilan University, Ramat-Gan, 52900, Israel.
[3]Sagol School of Neuroscience, Tel Aviv University, Tel Aviv, Israel.
[†]These authors contributed equally to this work.
[*]e-mail: ido.kanter@biu.ac.il

**Supplementary text:**

Reported phenomena are based on more than 50 examined cultures, where each type of experiment was repeated at least 10 times. Note that no hyperpolarization was observed in all examined neurons. The experimental measured RPs were found to cover a large range and in addition, reported range of the RPs are based on different types of neurons representing variety of spike waveforms. Results do not allow deducing the distribution of the RPs as well as preferred ranges. For instance, for experimental type 1 the ARP was found to be in the range of [2, 10] ms (experimental examples: 2, 3, 3.5, 4, 4.5, 5, 5.5, 6, 6.5, 7, 8, 8.3, 9, 10 ms), n=14, mean 5.8 ms and std=2.4 ms. For experimental type 2 the ARP was found to be in the range of [5, 30] ms (experimental examples: 5.5, 6, 6.5, 7, 8, 8.5, 9, 9.5, 10, 12, 12.5, 13, 13.5, 16, 21, 22, 23, 26 ms), n=18, mean 12.7 ms, std= 6.4 ms. For experimental type 3 the ARP was found to be in the range of [3, 13] ms (experimental examples: 3, 3.5, 4.5, 5.5, 6, 7.5, 8.5, 9, 10, 11, 12 ms), n=11, mean = 7.3 ms and std=3 ms.

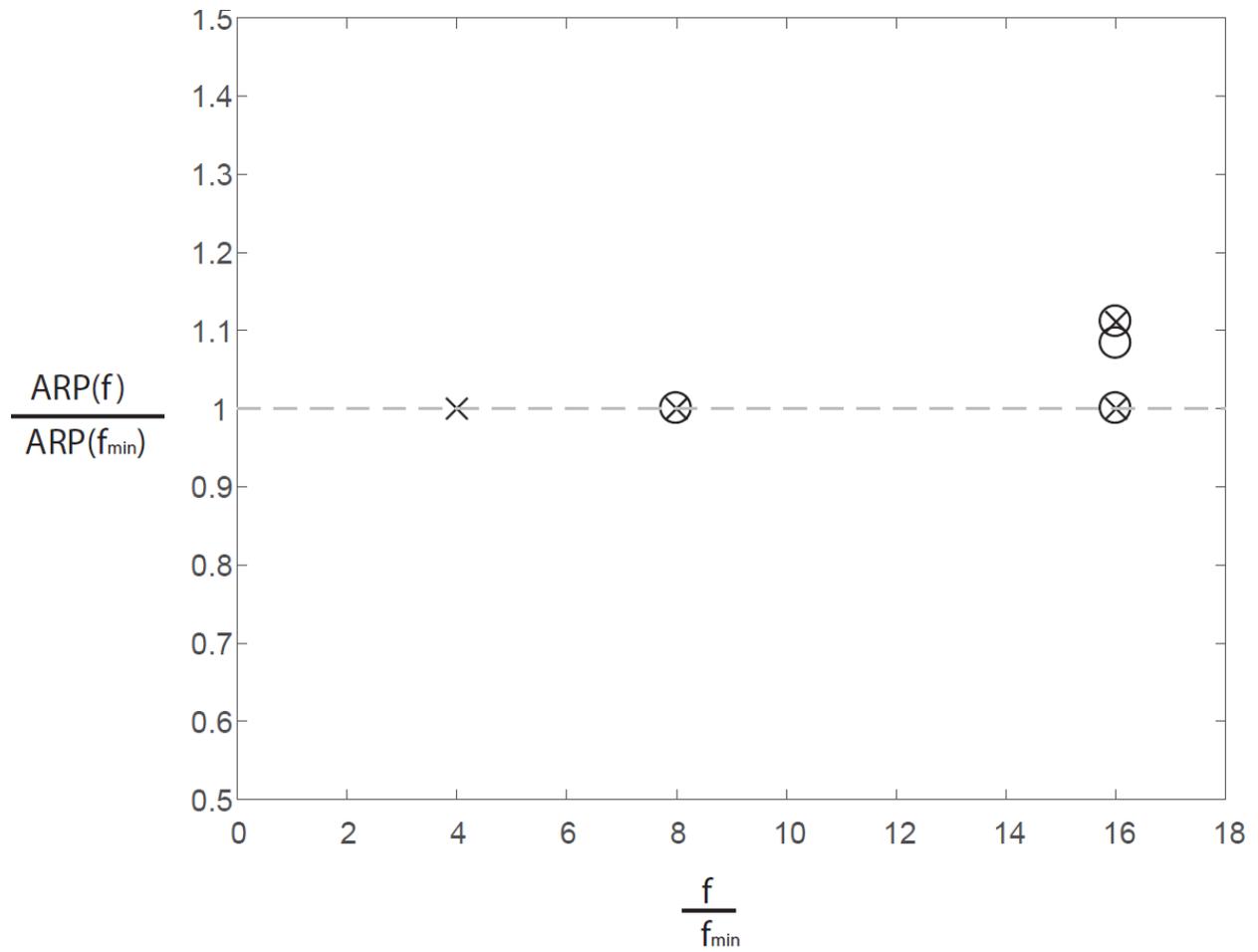

Fig. S1. Eight measurements of the ARP of different neurons at different frequencies (all below Fc), indicating that the ARP is robust to stimulation frequency (Fig. 3 in the manuscript).

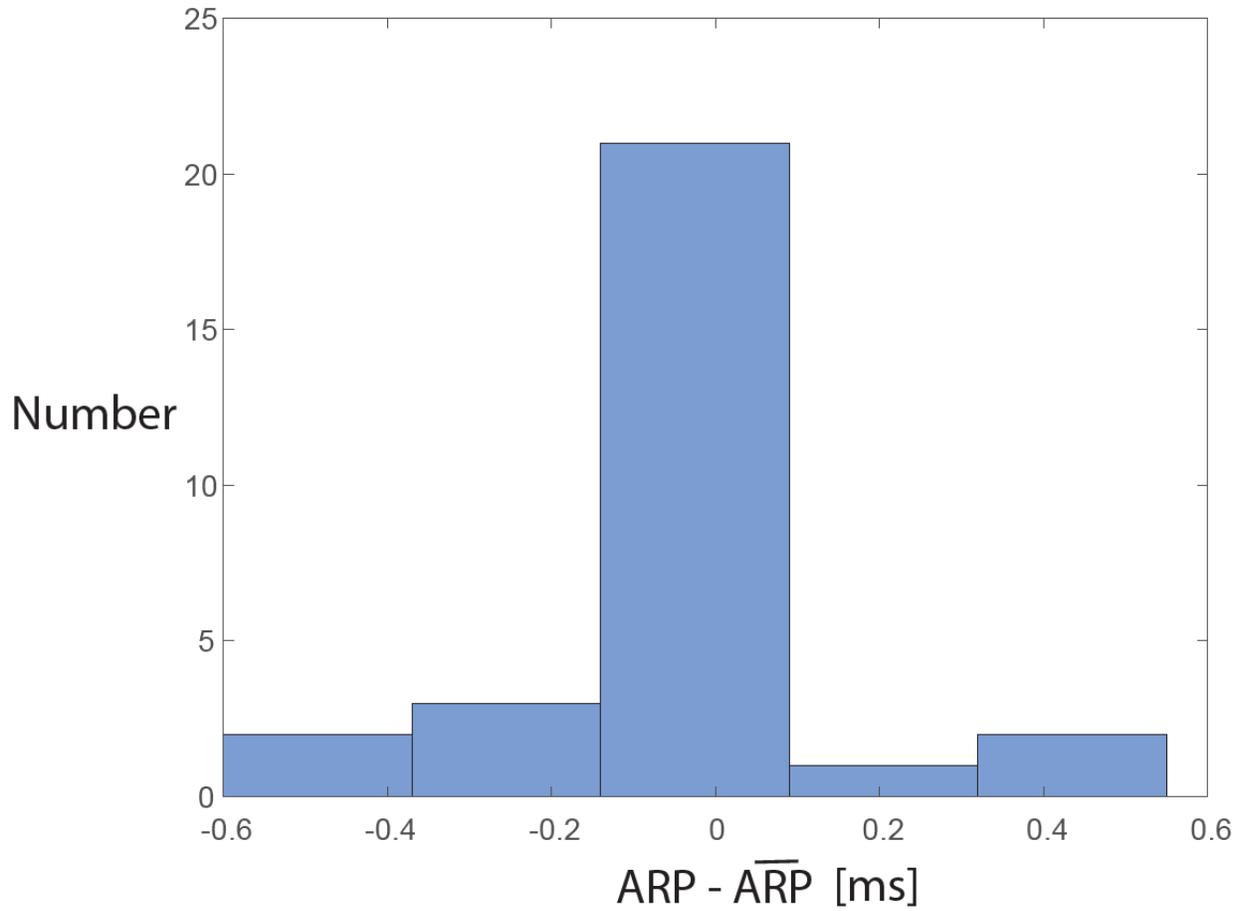

Fig. S2. A histogram of the deviations of repeated measurements of the ARP in each of eight neurons from their averaged value. Results indicate fluctuations below a millisecond, much below the anisotropic ARP effect (Figs. 4-5 in the manuscript).